\documentclass[12pt,a4paper,fleqn]{article}

\usepackage[english]{babel}
\usepackage[margin=25mm,headheight=15pt]{geometry}
\usepackage{fancyhdr}
\usepackage{indentfirst}
\usepackage{times}
\usepackage{titlesec}
\usepackage{hyperref}
\usepackage{soul}
\usepackage{caption}

\makeatletter
\@ifundefined{pdfoutput}% Definitely not using pdftex.
{% Standard TeX
\usepackage[dvips]{graphicx}
\usepackage[usenames,dvipsnames]{color}
}
{%  Running pdftex.
  \ifnum\pdfoutput=0\relax% Are we outputting pdf?
    \usepackage[dvips]{graphicx}
    \usepackage[usenames,dvipsnames]{color}
  \fi
  \ifnum\pdfoutput=1\relax% Are we outputting pdf?
    \usepackage[pdftex]{graphicx}
    \usepackage[usenames,dvipsnames]{color}
  \fi
}
\makeatother

\newcommand\cconfname{}
\newcommand\cconfdate{}
\newcommand\cconfmonth{}
\newcommand\cconfyear{}
\newcommand\cconfcity{}
\newcommand\cconfcountry{}
\newcommand\ccpapernum{}
\newcommand{\confname}[1]{\gdef\cconfname{#1}}
\newcommand{\confdate}[1]{\gdef\cconfdate{#1}}
\newcommand{\confmonth}[1]{\gdef\cconfmonth{#1}}
\newcommand{\confyear}[1]{\gdef\cconfyear{#1}}
\newcommand{\confcountry}[1]{\gdef\cconfcountry{#1}}
\newcommand{\confcity}[1]{\gdef\cconfcity{#1}}
\newcommand{\papernum}[1]{\gdef\ccpapernum{#1}}

\makeatletter
\def\@maketitle{%
  \vspace*{-8mm}%
  \vspace*{-5ex}%
  \thispagestyle{plain}%
  \begin{flushleft}%
    \hspace*{-0.7em}%
    \begin{tabular}{c}%
      \includegraphics[width=9.41cm]{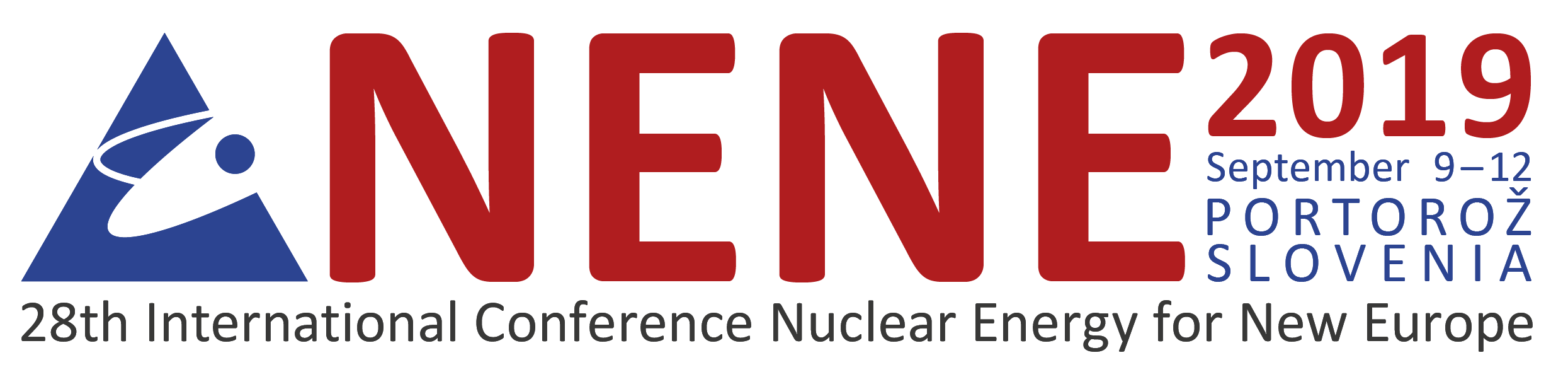}\\[4mm]%
    \end{tabular}%
  \end{flushleft}%
  \begin{center}%
    \textbf{ \textsf{\large \@title}}%
  \end{center}%
}
\makeatother

\newcommand{\Author}[2]{   \vspace*{-6mm}
                           \begin{center}
                           \textbf{#1} \par
                           {#2}
                           \end{center}
                       }

\titleformat{\section}{\normalsize\normalfont\bfseries}{\makebox[10mm][l]{\thesection}}{0pt}{\uppercase}
\titleformat{\subsection}{\normalsize\normalfont\bfseries}{\makebox[10mm][l]{\thesubsection}}{0pt}{}
\titleformat{\subsubsection}{\normalsize\normalfont\bfseries}{\makebox[10mm][l]{\thesubsubsection}}{0pt}{}

\makeatletter
\newcommand\footersize{\@setfontsize\footersize{9pt}{10}}
\makeatother
\fancypagestyle{plain}{%
\fancyhf{} % clear all header and footer fields
\fancyfoot[C]{\textcolor{Gray}{\footersize{\ccpapernum.\thepage}}}

}
\confname{Nuclear Energy for New Europe}
\confdate{9--12}
\confmonth{September}
\confyear{2019}
\confcity{Portoro\v{z}}
\confcountry{Slovenia}

\papernum{707}

\title{The initial step towards JOREK integration in IMAS}

\begin{document}
\maketitle

% !!!!!!!!!!!!!!!!!!!!!!!!!
% REPLACE WITH AUTHOR NAMES
% !!!!!!!!!!!!!!!!!!!!!!!!!
\Author{}{ \textbf{\underline{D. Penko}~$^1$, L. Kos~$^1$, G. Huijsmans~$^2$, S. D. Pinches~$^2$ }\\[1ex]
%, and EUROfusion IM Team~\footnote{See \href{http://www.euro-fusionscipub.org/eu-im}{http://www.euro-fusionscipub.org/eu-im}}}\\[1ex]
$^1$~Faculty of Mechanical Engineering, University of Ljubljana\\
A\v{s}kerčeva 6, 1000 Ljubljana, Slovenia\\
%leon.kos@lecad.fs.uni-lj.si, dejan.penko@lecad.fs.uni-lj.si\\[1ex]
$^2$~ITER Organization,
Route de Vinon-sur-Verdon - CS 90 046 - 13067 St Paul-lez-Durance Cedex - France}

%%%%%%%%%%%%%%%%%%%%%%%%%%%%%%%%%%%%%%%%%%%%%%%%%%%%%%%%%%%%%%%%%%%%%%
\section*{ABSTRACT}

JOREK~\cite{JOREK_bezier_surfaces, JOREK_website} is a non-linear magnetohydrodynamic (MHD) code which was developed with the intent of producing simulations of the MHD instabilities occurring in magnetically confined plasmas. Such simulations substantially contribute to the understanding of MHD instabilities such as Edge Localised Modes (ELMs) and are essential for the optimization of future fusion devices such as ITER. The code itself is already well established and has been validated on many occasions through simulations of MHD instabilities related to present fusion devices JET, MAST, ASDEX Upgrade, and DIII-D.

% The code is based on robust fully implicit numerics, and takes the full toroidal X-point geometry consisting of core, scrape-off layer, divertor region, and separatrix. 

%The MHD instabilities, of which understanding is essential for the optimization of the future fusion devices such as ITER, are edge localised modes (ELMs) and disruptions such as massive gas injection, shattered pellets, vertical displacement events (VDEs), runaway electrons, tearing mode seeding and suppression

JOREK is being adapted to work with the Integrated Modelling \& Analysis Suite (IMAS) which is being actively developed and used by the ITER Organization, the EUROfusion community and other ITER Members. The list of codes adapted to use the IMAS Data Model is gradually increasing with examples including SOLPS-ITER~\cite{presentation_of_solpsiter} and JINTRAC~\cite{JINTRAC}. The main goal of the integration of JOREK with IMAS is to enable interaction with the plasma scenarios stored in the IMAS databases in the form of Interface Data Structures (IDSs): input conditions can be read from the databases and nonlinear plasma states determined by JOREK stored.  IDSs provide a uniform way of representing data within the IMAS framework and allow to transfer data between codes and to storage within larger integrated modelling workflows. In order to integrate JOREK within IMAS it is therefore necessary that transformation tools are developed to facilitate the reading and writing of the relevant IDSs, including the \emph{MHD} IDS, with its underlying \emph{Generalized Grid Description} (GGD). For this purpose, utilities have been developed that extract JOREK simulation plasma state, namely the grid geometry and computed physical quantities for each time slice, and then transform them to the appropriate output IDSs. In this article, these initial steps towards full JOREK integration into IMAS will be presented.

%%%%%%%%%%%%%%%%%%%%%%%%%%%%%%%%%%%%%%%%%%%%%%%%%%%%%%%%%%%%%%%%%%%%%%
\clearpage
\section{INTRODUCTION}

Magnetohydrodynamic (MHD) instabilities are one of the key factors in the optimization of the ITER machine operation. In order to either avoid, mitigate or at least control such instabilities is crucial to understand the physics behind them, their effects, and to identify what is causing them. A few of the significant MHD instabilities occurring at fast time scales in tokamak plasmas are:
\begin{itemize}
    \item disturbances, which can lead to rapid losses of plasma energy and density and can consequently lead to fast plasma termination, and
    \item Edge Localized Modes (ELMs)~\cite{elms1}, periodic disturbances occurring in tokamaks resulting in a fraction of the plasma energy present in the confined hot edge plasma being transferred to the tokamak wall, potentially causing heat loads, surface temperature increase and subsequently increasing the risk of damaging the Plasma Facing Components (PFC).
\end{itemize}

Since extrapolation of plasma behaviour and MHD control requirements from current experiments to ITER is not straightforward and because the ITER plasma parameters cannot be obtained in the present tokamak devices, the simulations for predicting the MHD instabilities are essential. There are quite a few plasma codes in existence that deal with such problems. One of such codes is JOREK~\cite{JOREK_bezier_surfaces, JOREK_website}, a well established  non-linear MHD code providing the means for producing simulations of the MHD instabilities occurring in magnetically confined plasmas such as ITER.

With the intent to enhance and support the ITER research activities, the Integrated Modelling \& Analysis Suite (IMAS)~\cite{imbeaux15:_desig_iter, _IMAS_implementation} is being progressively established as the main framework for research activities on ITER tokamak experiment. Presently the suite is being actively used by the ITER Organisation, the EUROfusion community, and other ITER collaborators. The JOREK code is yet to be involved in ITER integrated modelling campaign and in order to bring JOREK closer to this use its integration into IMAS is required. The initial ambition of JOREK integration within IMAS is to enable interaction with the plasma scenarios stored in the IMAS databases in the form of Interface Data Structures (IDSs)~\cite{IDS_data_dictionary, IMAS4SOLPS-ITER, EPS_IMAS_ITER_wall}. For that purposes, it is essential that suitable transformation tools are developed to facilitate the reading and writing operations for the relevant IDSs and its substructures. 

\section{IMAS STATUS}

The ITER Integrated Modelling \& Analysis Suite (IMAS)~\cite{imbeaux15:_desig_iter, _IMAS_implementation} is a scientific software infrastructure providing foundations for collective development and execution of integrated plasma applications and plasma codes describing fusion operations in tokamak experiments such as ITER. The collection of plasma codes integrated within IMAS Data Model is gradually increasing. A few of such codes are SOLPS-ITER~\cite{presentation_of_solpsiter, IMAS4SOLPS-ITER} and  JINTRAC~\cite{JINTRAC}. More of such codes are planned to be integrated in IMAS including JOREK. Even though an applicable IMAS is already available it is being constantly improved and further developed by the ITER community. Presently, the IMAS is mainly available for users on ITER GPC and EUROfusion Gateway HPC. 

\begin{figure}[htbp]
    \centering
    \captionsetup{justification=centering}
    \includegraphics
    [width=0.9\linewidth]
    {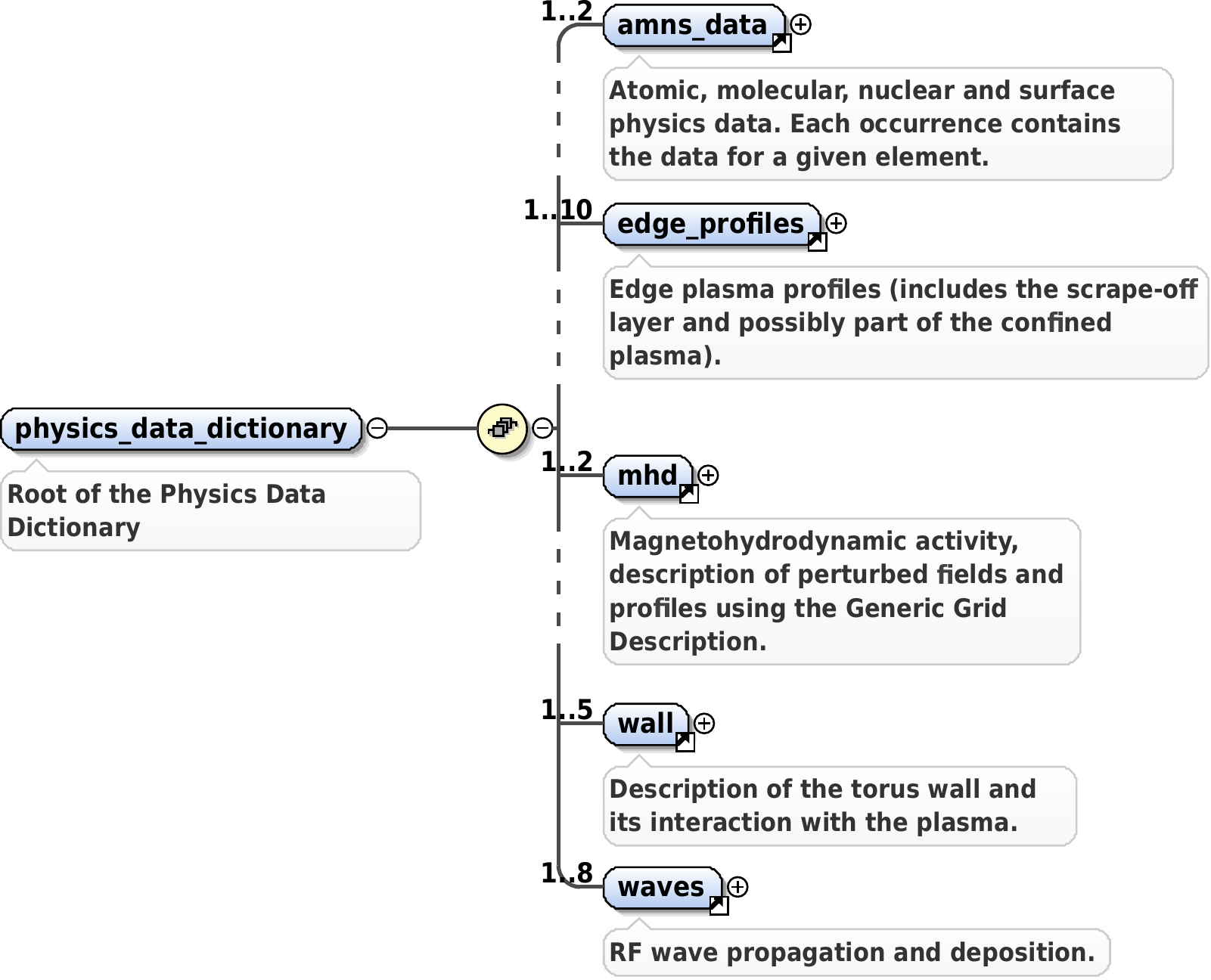}
    \caption
    {The presentation of Data Dictionary root node displaying a few of the IDSs. Presently, the Data Dictionary consists of more than 50 different IDSs that are constantly being improved and updated. The IDS relevant to the non-linear MHD data is the \textit{MHD} IDS.}
    \label{fig:dd_schema_root}
\end{figure}

The IMAS suite is based on an underlying \textit{Physics Data Model} (PDM) which represent the basis of coupling the plasma codes with standardized database structured named \emph{Interface Data Structures (IDSs)}~\cite{IDS_data_dictionary, IMAS4SOLPS-ITER, EPS_IMAS_ITER_wall}. IDSs are a large-scale and complex standardized tree-like hierarchical segments of a \textit{Data Dictionary}, an extensive database description connecting all IDSs into a single structured data composition, as shown in Fig.~\ref{fig:dd_schema_root}. IDSs feature standardized data archival and retrieval in this way unifying the fusion data format, contributing to more straightforward data interpretation, data comparison, and case repeatability. Furthermore, these databases are accessible from personal or global archives on the same HPC cluster with the use of \textit{MDSplus}~\cite{mdsplus_web_intro}, a set of software tools for data management of complex scientific data, ensuring efficient data tracing. IDSs are translated from the XSD schema and are accessible with several programming languages (Fortran, C++, Java, Python, and Matlab).

\begin{figure}[htbp]
    \centering
    \captionsetup{justification=centering}
    \includegraphics
    [width=0.73\linewidth, trim={0px 0px 0px 0px},clip]
    {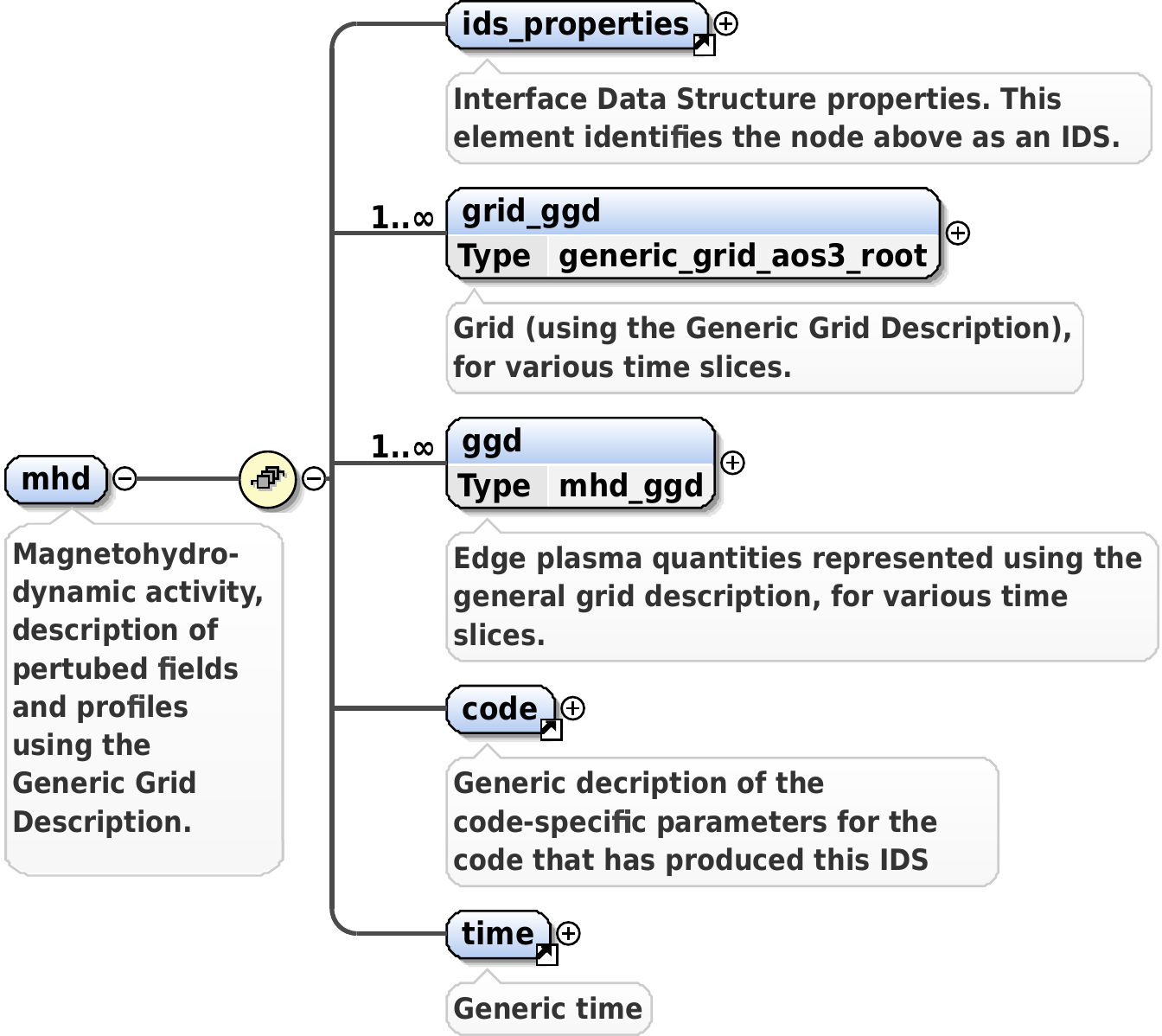}
    \caption
    {An overview of the top level of the MHD IDS tree structure and its substructures \texttt{ids\_properties}, \texttt{grid\_ggd}, \texttt{ggd}, \texttt{code} and \texttt{time}.}
    \label{fig:dd_schema_mhd}
\end{figure}

Each IDS is set to contain a wide amount of information corresponding to a specific field relevant to plasma state and fusion device operations such as the tokamak and its components geometry description, plasma equilibrium, diagnostics, heating and fuelling sources, Scrape-Off Layer (SOL) plasma state description, control systems, etc. The IDSs relevant to the MHD data are \textit{MHD} and \textit{MHD Linear} IDSs. The stored data in IDSs can then be accessed and further processed in many post-process applications. One of the important central IDS substructures is the General Grid Description (GGD). GGD as a term encompass two separate IDS substructures: \texttt{ggd} and \texttt{grid\_ggd} substructure, as shown in Fig.\ref{fig:dd_schema_mhd}. The \texttt{ggd\_grid} is an array of structures designed to store all n-dimensional grid geometry description data regarding the tokamak regions while the \texttt{ggd} node is an array of structures designed to store all plasma state quantities under its own timebase. \textit{MHD linear} IDS does not contain the GGD structures and it is set to contain linear MHD information while \textit{MHD} IDS does contain the required GGD structures and it contains data substructures suitable for non-linear MHD data. Due to that, the work presented in this paper is focused only on the \textit{MHD} IDS. An overview of the \emph{MHD} IDS structure is shown in Fig.~\ref{fig:dd_schema_mhd}. 

\subsection{DEVELOPED IMAS UTILITIES FOR JOREK}

    The JOREK computation results are by default stored in the form of a multiple \emph{Hierarchical Data Format} (HDF5) files with each HDF5 file containing the nonlinear plasma state of a single time slice. The total number of time slices and therefore HDF5 files depends on the pre-set run parameter \texttt{nstep} which defines the number of time step computations to be done by JOREK. The time step is defined by parameter \texttt{tstep}. The JOREK project GIT repository on ITER~\cite{JOREK_git} contains, besides the JOREK source code, also a set of post-processing tools and utilities which allow data transformation and visualization of the HDF5 files.
    A few of such utilities, which in addition provide a good insight in the contents of the HDF5 files, are:
    \begin{itemize}
        \item \texttt{jorek2vtk}, a utility allowing data format conversion of a single HDF5 file to Visualization Toolkit (VTK) file format,
        \item \texttt{jorek\_read\_h5.py} providing Python routines for reading the contents of the HDF5 files, and
        \item \texttt{JOREK ParaView plugin} which allows reading the MHD data from all HDF5 files - time slices at once and passing it to ParaView~\cite{paraview_guide, paraview_webpage} enabling all ParaView built-in features for data post-processing and data analysis to be used on the computed JOREK plasma state. One of the convenient features as also a time slice browser and the option to run a sequence of plasma state time slices as an animation-like visualization.
\end{itemize}

%With the idea to store the JOREK plasma state output to the MHD IDS and with no known applications available a new set of utilities for that purpose needed to be developed. 

With the initial focus on converted VTK files and later on base HDF5 files, a set of modules and utilities have been developed in the Python3 programming language with the intent to provide the basis of the JOREK involvement in IMAS:

\begin{itemize}
    \item \texttt{jorekVTKtoIDS}, a tool allowing extracting the plasma state from a single VTK file (single time slice) and writing it to specified \textit{MHD} IDS, and
    \item \texttt{jorekHDF5toIDS}, a tool that extracts the plasma state from all available HDF5 files (representing data for all time slices) produced by JOREK computation and writes the whole dataset to a single \emph{MHD} IDSs. As the grid geometry does not change with time in the JOREK simulations, it is extracted from the first time step and written only to a single \texttt{grid\_ggd[:]} structure, as re-writing it for each time-slice is redundant. The plasma state physical quantities are written for each time slice to its separate \texttt{ggd[time\_slice]} structure where \texttt{time\_slice} represents a time slice index. In this way the plasma state of each time slice is being written to GGD. 
\end{itemize}

The above-listed tools write the extracted grid geometry and the following computed physical quantities to the \textit{MHD} IDS:

\begin{itemize}
    \item poloidal magnetic flux $\psi$,
    \item electric potential $\phi$,
    \item parallel velocity $\vec{v}_\parallel$,
    \item toroidal current density $\vec{j}_{tor}$,
    \item total temperature $T$, 
    \item vorticity $\omega$, and
    \item mass density $\rho$.
\end{itemize}

%The remaining quantities are:
%\begin{itemize}
%    \item $\rho$ (density),
%\end{itemize}

\begin{figure}[htbp]
    \centering
    \captionsetup{justification=centering}
    \includegraphics
    [width=0.90\linewidth]
    {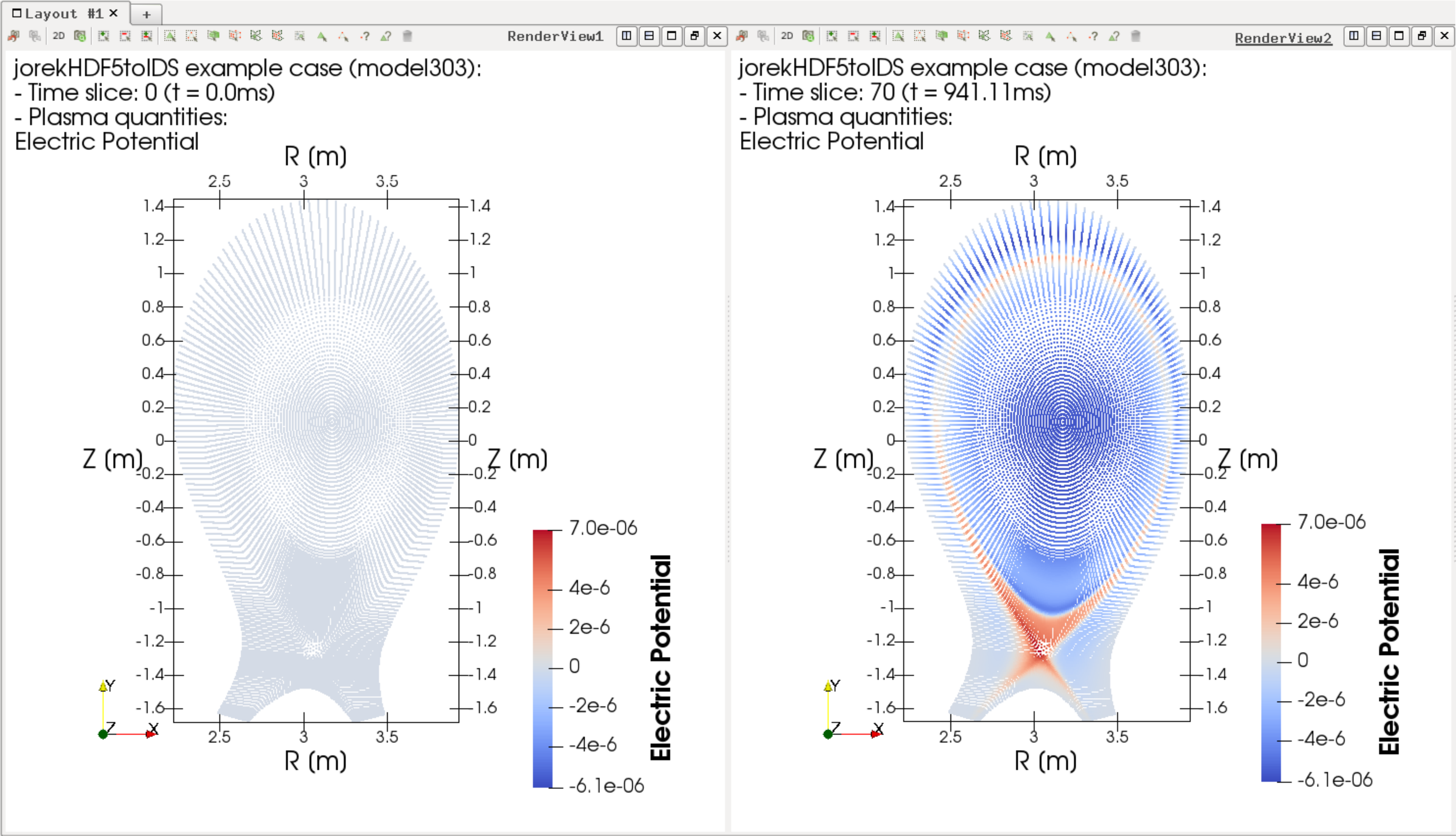}
    \caption
    {\emph{ReadUALEdge}: An example of dat acomparison of Electric Potential between the first (t = 0.0ms) and 71st time slice (t = 941.11ms) obtained by the JOREK \texttt{model303} computation. The integration in ParaView enables the display of the passed data and the use of ParaView built-in features for data analysis. }
    \label{fig:0vs70}
\end{figure}

\begin{figure}[htbp]
    \centering
    \captionsetup{justification=centering}
    \includegraphics
    [width=0.90\linewidth]
    {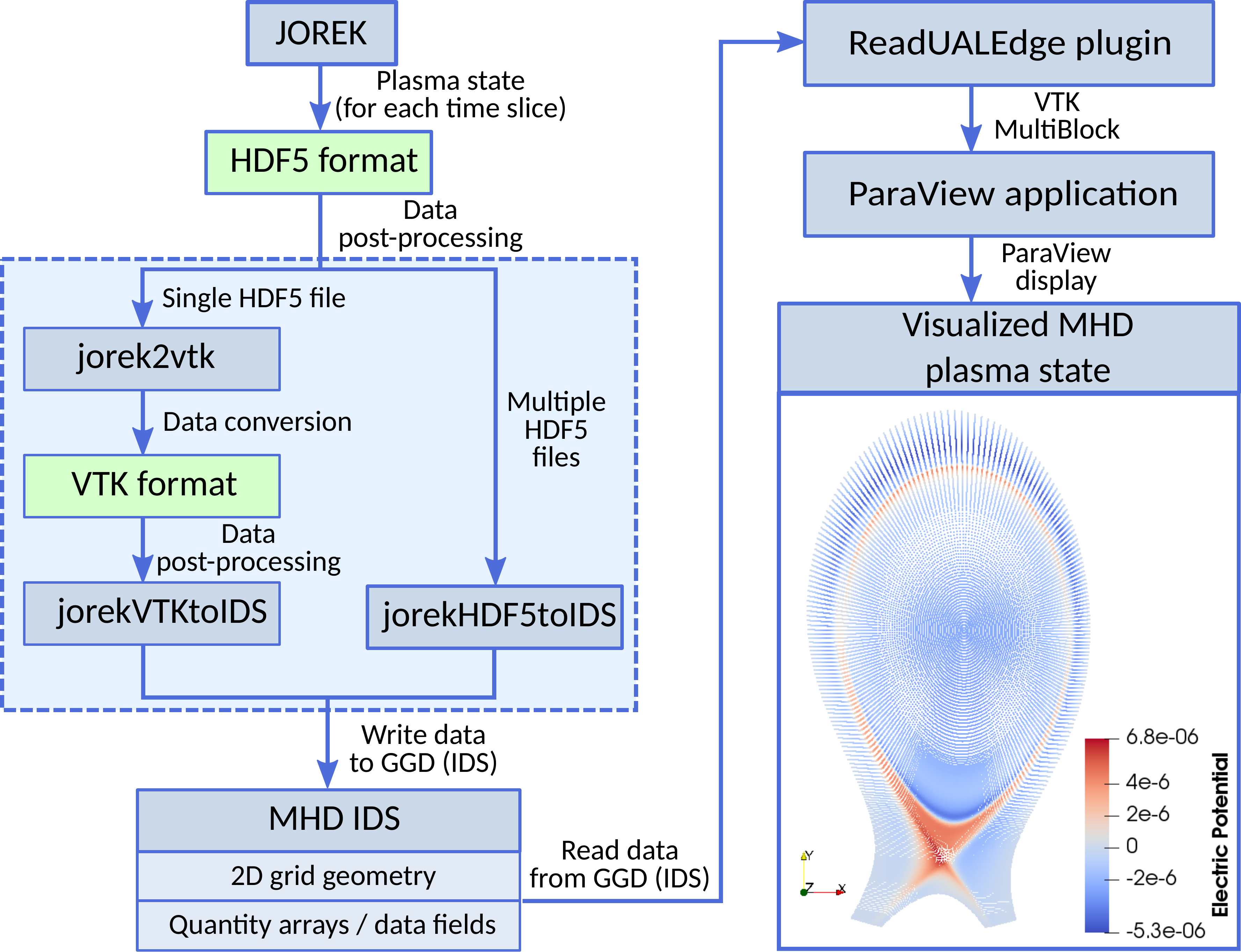}
    \caption
    {Presentation of the complete workflow from JOREK computation to visualization in ParaView.}
    \label{fig:workflow}
\end{figure}

In order to read the data directly from the produced IDS and to visualize its contents a suitable tool was required. One of the accessible and applicable tools for the visualization of the data from the IDSs is the \emph{ReadUALEdge} plugin, a component of the \emph{SOLPS-GUI} framework~\cite{IMAS4SOLPS-ITER, solpsgui_git}. The \textit{ReadUALEdge} ParaView plugin was initially intended for passing the data from \textit{Edge Profiles}, \textit{Edge Transport} and \textit{Edge Sources} IDSs to ParaView. Presently, there are two codes that write their plasma state output to those IDSs: SOLPS-ITER and EDGE2D (JINTRAC component). The plugin itself provides a suitable foundation for extracting the data from the IDSs GGD structure for any time slice. For this reason, the plugin was further extended and improved allowing the same functionality for MHD IDS, as shown in Fig.~\ref{fig:0vs70}. The full workflow from JOREK case computation to ParaView visualization is shown in Fig.~\ref{fig:workflow}.

\clearpage
\section{CONCLUSION}

The additions presented in this paper provide the first aspects and methods to support the future activities and developments towards the final goal of full JOREK code integration in IMAS, consequently supporting the plasma research activities and contributing to overall IMAS versatility. 

The developed post-processing utilities provide a convenient way of writing the JOREK computation plasma state to the standardized IDSs and visualizing the data stored in them. The IDSs GGD structure proved as a valuable asset for the integration of JOREK since it allows storage of multiple plasma state time slices in a single IDS while leaving out the duplicate data. Moreover, the \textit{ReadUALEdge} plugin proved to be a decent foundation for visualizing any spatial data stored in the GGD structure. The plugin could be further improved by enabling the display of 3D data and to mimic the \texttt{JOREK ParaView plugin} feature which allows loading multiple time slices at once and running them in a sequence resulting in a convenient presentation of plasma behaviour in tokamak devices. 

The presented developments have the potential to be further improved and they provide the initial basis for the planned JOREK and IMAS related activities. There are more steps necessary to be done before JOREK integration in IMAS is finished. Firstly, writing the remaining variables fully describing the plasma state and JOREK case run description to the MHD IDS is required. Next, it is necessary to enable IDSs to be used as an optional input format for JOREK computation in which case the input IDS would require to hold all necessary case input conditions and the input parameters. Finally, an optional feature should be introduced to JOREK which would allow writing the computed plasma state directly to MHD IDS. With those major JOREK pre- and post-processing operations the integration of JOREK in IMAS would be considerably closer to completion. 

\section*{ACKNOWLEDGMENTS}

This work has been carried out in collaboration with the ITER Organization. The initial version of the SOLPS-GUI ReadUALEdge plugin was partially conducted under the ITER  contract  IO/4300001173. The views and opinions expressed herein do not necessarily reflect those of the ITER Organization.

% Redefine the references label to be in uppercase
\renewcommand*{\refname}{\normalfont\bfseries\uppercase{REFERENCES}}
\renewcommand{\arraystretch}{0.1}
\setlength{\parsep}{0.1cm} \setlength{\itemsep}{0.1cm}
\bibliographystyle{unsrt}
%\bibliography{References}

\addcontentsline{}{chapter}{Bibliography}

\end{document}